\documentclass[11pt]{pazh} 

\usepackage[T2A]{fontenc}
\usepackage{graphics}
\usepackage{graphicx}
\usepackage{latexsym}
\usepackage{amssymb}
\usepackage{amsmath}
\usepackage{hyperref}
\usepackage{soul}

\begin{document}

\title{\bf Active galactic nuclei among polar-ring galaxies}

\author{D.V. Smirnov\inst{1}, V.P. Reshetnikov\inst{1}
}

\institute{ $^1$ St.Petersburg State University, Universitetskii pr. 28, St.Petersburg, 
198504 Russia
}

\abstract{
Based on SDSS data, we consider the fraction of active galactic nuclei among 
polar-ring galaxies. We have found evidence for an excess of Seyfert galaxies 
and LINERs among polar-ring galaxies compared to ordinary objects. 
The observed nuclear activity of polar-ring galaxies may be associated with
the accretion of gas from the region of the polar structures onto the central 
galaxies.
}
\titlerunning{Active nuclei among PRGs}
\maketitle

%%%%%%%%%%%%%%%%%%%%%%%%%%%%%%%%%%%%%%%%%%%%%%%%%%%%%%%%%%%%%%%%%%%%
\section{Introduction}
%%%%%%%%%%%%%%%%%%%%%%%%%%%%%%%%%%%%%%%%%%%%%%%%%%%%%%%%%%%%%%%%%%%%

Polar-ring galaxies (PRGs) are a very rare class of extragalactic objects. 
PRGs consist of a central galaxy surrounded by an extended ring or disk 
approximately along its minor axis. (In what follows, we will use the term 
`PRG' to designate a class of objects with circumpolar optical structures without 
without dividing into polar-disc or polar-ring galaxies.)
Numerous examples of such systems are given in the Polar-Ring 
Catalogue (PRC, Whitmore et al. 1990) and the SDSS-based Polar Ring 
Catalogue (SPRC, Moiseev et al. 2011) and the paper by Reshetnikov and 
Mosenkov (2019) (hereafter RM2019).

The central objects in most PRGs are gas-poor early-type (E/S0) galaxies 
(see, e.g., Whitmore et al. 1987; Finkelman et al. 2012). The polar
structures show rotation around the major axes of the central galaxies, 
they contain stars and gas, they have relatively blue colors, there is evidence
of ongoing star formation, and the gas in the polar structures often has a 
subsolar metallicity (Whitmore et al. 1987; Reshetnikov and Combes 1994, 2015;
Egorov and Moiseev 2019; etc.). The relative fraction of PRGs among nearby 
galaxies is $\sim$10$^{-3}$ (Reshetnikov et al. 2011) and they avoid a dense 
spatial environment (Finkelman et al. 2012; Savchenko and Reshetnikov 2017).

To explain the presence of two morphologically and kinematically decoupled 
large-scale subsystems in one object, it is most often assumed that PRGs
were formed in two steps. First the central galaxy was formed and then the 
polar structure arose from some ``secondary'' event. This secondary event could
be, for example, the merging of galaxies with mutually orthogonal disks, 
the capture of a companion into a circumpolar orbit and its disruption, 
the capture of matter from an approached galaxy, and the accretion of matter 
from intergalactic space (see, e.g., Bekki 1997; Reshetnikov and Sotnikova 1997; 
Bournaud and Combes 2003; Maccio et al. 2006; Brook et al. 2008).

The processes listed above could leave certain traces on the periphery of 
galaxies in the form of faint extended structures (shells, tidal tails). On the
other hand, the formation of polar rings should have been accompanied by the 
accretion of matter onto the central galaxy. For example, numerical simulations
of the PRG formation during the merger of elliptical and spiral galaxies show 
that 10-20\% of the gas falls to the center of the formed PRG (Bournaud and
Combes 2003). The evolution of the polar rings, given their interaction with 
the main galaxies, can also lead to the loss of angular momentum of the gas and its
accretion to the center (see, e.g., Wakamatsu 1993).

The infall of matter onto the massive central object at the centers of galaxies 
is the universally accepted mechanism of nonthermal activity of their 
nuclei. Therefore, given the features of formation and evolution of PRGs, 
the question about the fraction of active galactic nuclei (AGNs) among PRGs
arises. This question still remains almost unexplored (Reshetnikov et al. 2001; 
Finkelman et al. 2012), which, at least in part, is related to the small number
of known PRGs. New samples of PRG candidates selected from 
SDSS\footnote{https://www.sdss.org} (SPRC; RM2019) 
have appeared in recent years, which allows this problem to be considered based 
on more data that have been available previously.

Throughout this article, we adopt a standard flat CDM cosmology with
$\Omega_m=0.3$, $\Omega_{\Lambda}=0.7$, $H_0 = 70$ km s$^{-1}$ Mpc$^{-1}$.

%%%%%%%%%%%%%%%%%%%%%%%%%%%%%%%%%%%%%%%%%%%%%%%%%%%%%%%%%%%%%%%%%%%%
\section{Sample of galaxies}
%%%%%%%%%%%%%%%%%%%%%%%%%%%%%%%%%%%%%%%%%%%%%%%%%%%%%%%%%%%%%%%%%%%%

To study the nuclear activity of PRGs, we considered the objects from two 
works based on SDSS. Most of the galaxies in our sample were taken from
the SPRC (Moiseev et al. 2011), which provides data for 275 PRG candidates. 
These candidates were divided into 4 groups; 70 and 115 galaxies
were attributed to the best and good PRG candidates, respectively. 
The third group includes 53 objects possibly related to PRGs (galaxies 
with strong warps of their stellar disks, interacting and merging
galaxies). The galaxy SPRC-198 from this group turned out to duplicate the 
galaxy SPRC-102 and, therefore, we left 52 objects in the subclass of 
PRG-related objects. Consequently, the total number of original objects in the 
SPRC is 274. The fourth group includes 37 candidates in which the presumed
polar structures may be seen face-on. Spectroscopic observations of SPRC 
objects show that the bulk of the first group (best candidates) are kinematically
confirmed PRGs (Moiseev et al. 2011, 2015; Egorov and Moiseev 2019). 
RM2019 may be considered as a supplement to the SPRC -- 31 galaxies from SDSS
morphologically similar to the best candidates in the SPRC are described in 
this paper.

%%%%%%%%%%%%%%%%%%%%%%%%%%%%%%%%%%%%%%%%%%%%%%%%%%%%%%%%%%%%%%%%%%%%

\begin{table*}
\caption{The samples of PRG candidates}
\begin{center}
\begin{tabular}{c|c|c|c|c}
\hline
Sample &  Number of PRGs &  $\langle M_r \rangle$ & $\langle g-r \rangle$ & Final number of PRGs\\
\hline          
SPRC (all)        & 274 &  -21.10$\pm$1.21  & +0.65$\pm$0.21 & 176 \\ % 265
Best candidates  & 70  &  -21.20$\pm$0.77  & +0.72$\pm$0.13 & 45 \\ % 66
Good candidates  & 115 &  -20.79$\pm$1.36  & +0.58$\pm$0.19 & 81 \\ % 111
Related objects  & 52  &  -21.43$\pm$1.15  & +0.71$\pm$0.28 & 29 \\ % 51
Face-on rings    & 37  &  -21.36$\pm$1.20  & +0.66$\pm$0.23 & 21 \\ % 37
RM2019          & 31  & -20.97$\pm$1.01 & +0.70$\pm$0.14     & 7  \\
\hline
\end{tabular}
\end{center}
\end{table*}
%%%%%%%%%%%%%%%%%%%%%%%%%%%%%%%%%%%%%%%%%%%%%%%%%%%%%%%%%%%%%%%%%%%%

Table\,1 summarizes the main characteristics of the samples under study. 
Columns 2, 3, and 4 in this table give, respectively, the number of objects 
in the corresponding sample, the mean $r$-band absolute magnitudes of
the galaxies, and the mean $g - r$ colors. The absolute magnitudes and colors 
were corrected for Galactic extinction (Schlafly and Finkbeiner 2011) and the 
$k$-correction (Chilingarian et al. 2010).

Then, for the PRG samples we extracted the fluxes for the emission lines used 
in the classification (see the next section) from SDSS using the TOPCAT
package (Taylor 2005). Only those galaxies for which the signal-to-noise ratio 
for all emission lines exceeded 3 were left in the samples (the same condition
was also used to produce the comparison samples -- see below). This constraint 
reduced the sizes of the original samples. The last column in Table\,1 gives the
final numbers of objects used to classify the types of nuclear activity.

Based on SDSS data (DR15, Aguado et al. 2019), we constructed the comparison 
samples for each of the PRG subgroups in Table\,1. The samples were
produced in such a way that the distributions of objects in them in luminosity, 
$g - r$ color, and redshift were close to the observed distributions for PRGs 
(see an example in Fig.\,1). For this purpose, galaxies were
randomly extracted from SDSS in accordance with the probability densities 
specified by the observed distributions of PRG characteristics. The sizes of the
comparison samples vary from $\sim$17000 for the PRG-related objects to 
55000 for the total PRG sample.

We consider the PRGs as single objects without separating the contribution of 
the central galaxy. If this is done, then at typical PRG parameters 
(Reshetnikov and Combes 2015) the corrections will be relatively small -- 
we should reduce the $r$-band absolute luminosities of the galaxies, on average, 
by 0.$^m$3 and increase the $g - r$ color by 0.$^m$03.

\begin{figure}
\centering
\includegraphics[width=9.3cm, angle=0, clip=]{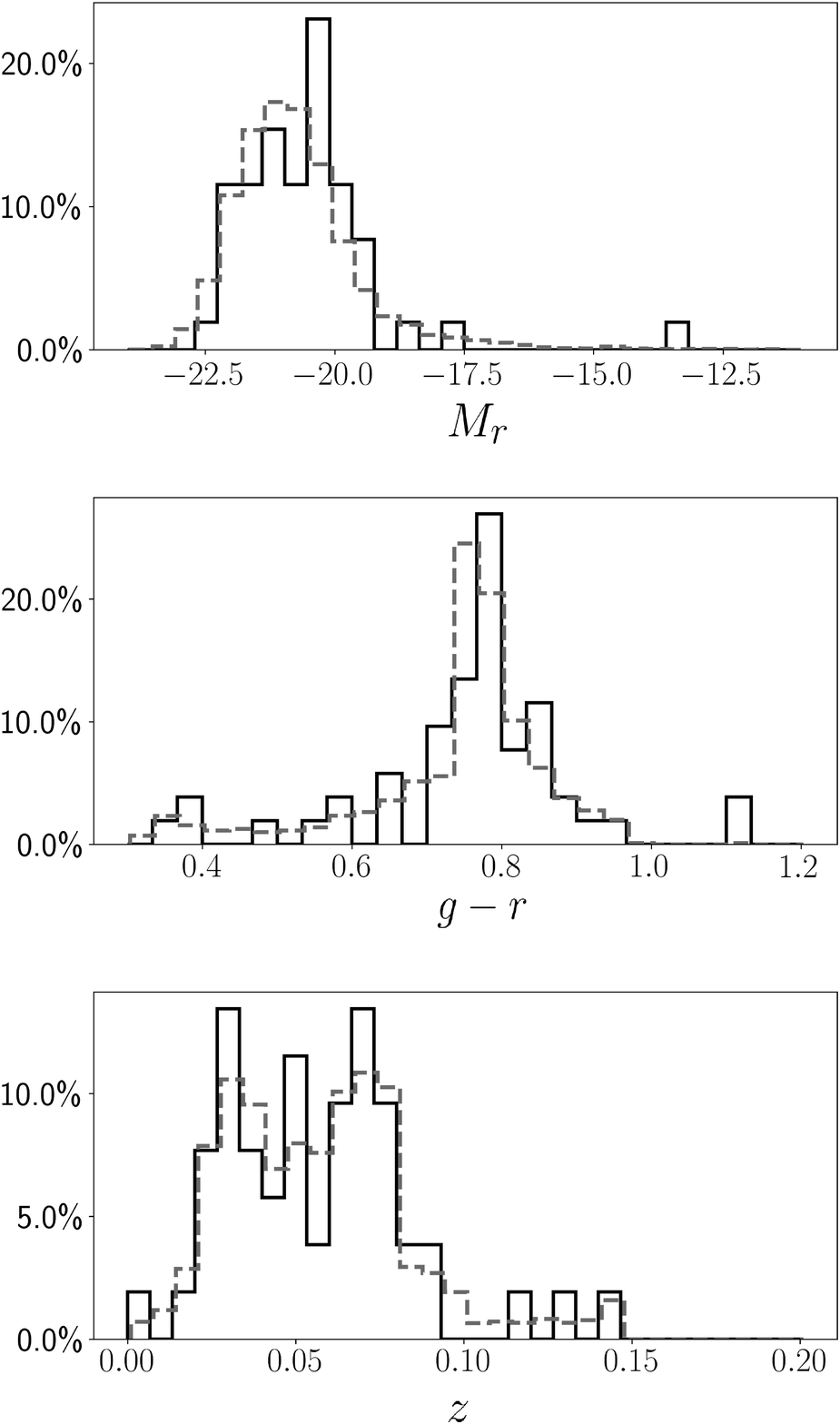}\\
\caption{Distributions of the best PRG candidates (with the addition of the 
objects from RM2019) in absolute magnitude, $g - r$ color, and redshift 
(solid line). The dashed line indicates the corresponding distributions 
for the comparison sample.}
\end{figure}

\section{Classification of galaxies}

To classify the types of PRG nuclei, we used the classical BPT diagrams 
(Baldwin et al. 1981) extended later by other authors (see, e.g., Veilleux
and Osterbrock 1987; Kewley et al. 2001). The emission line ratios 
[OIII]$\lambda$5007/H$\beta$, [NII]$\lambda$6583/H$\alpha$,
[SII]/H$\alpha$ (here and below, [SII] denotes the sum of
the lines, [SII]$\lambda$6717+[SII]$\lambda$6731), and [OI]$\lambda$6300/H$\alpha$
are compared on these diagrams to separate the objects with different 
ionization sources. We did not include the 
[OIII]$\lambda$5007/[OII]$\lambda$3727 -- [OI]/H$\alpha$ diagram in our 
analysis, because it does not separate the star-forming galaxies and 
galaxies with composite nuclei (see below). Furthermore, its inclusion reduces
the PRG sample. The observed line intensities for the PRGs and the objects 
in the comparison samples were corrected for extinction using the Balmer
decrement (the true ratio $I(H\alpha)/I(H\beta)$ was taken to
be 2.86; see Osterbrock and Ferland 2006) and the interstellar extinction 
curve from Calzetti (1997).

%%%%%%%%%%%%%%%%%%%%%%%%%%%%%%%%%%%%%%%%%%%%%%%%%%%%%%%%%%%%%%%%%%%%
\begin{figure}
\centering
\includegraphics[width=7.5cm, clip=]{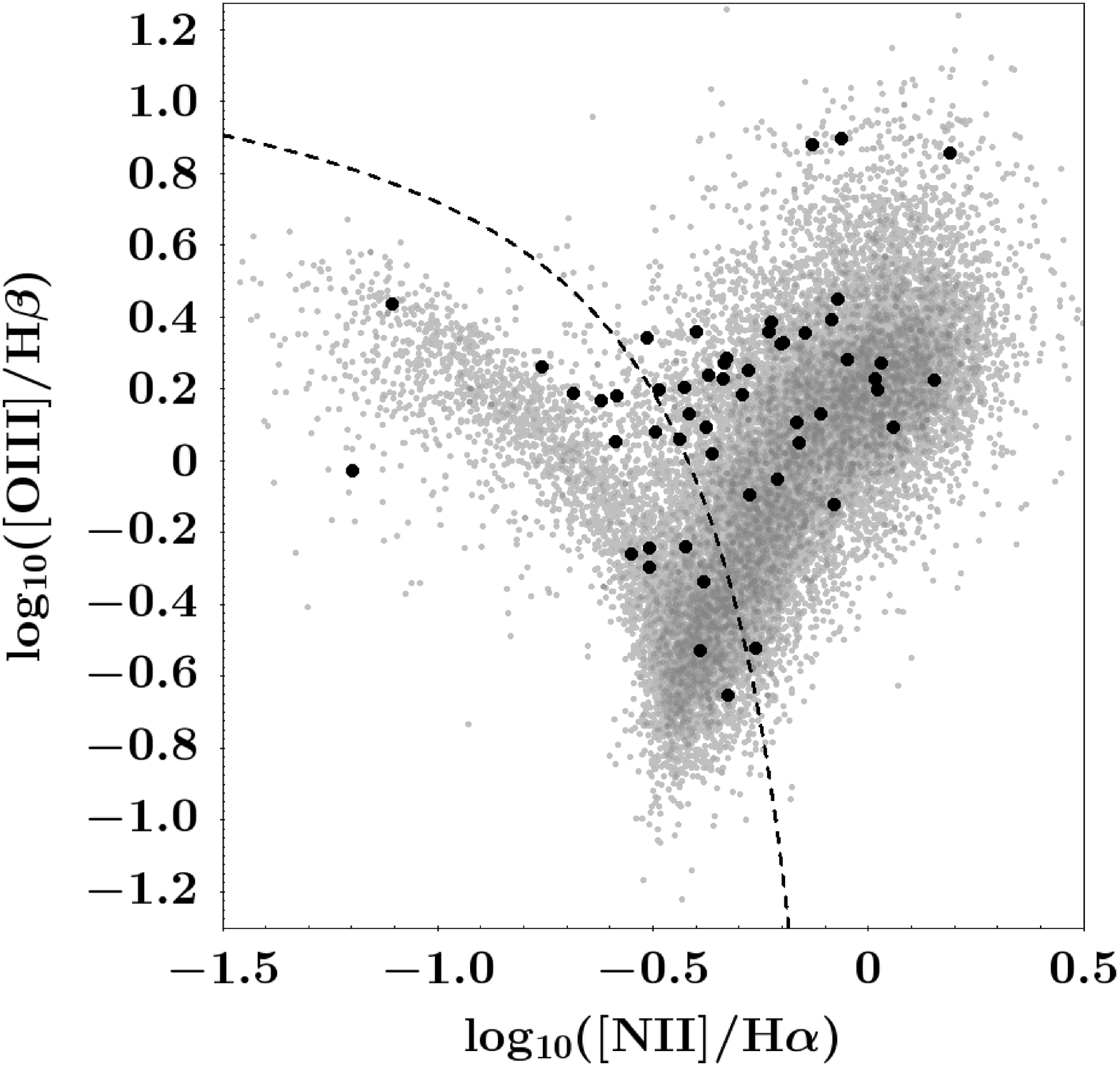}\\

\vspace{0.4cm}

\includegraphics[width=7.5cm, clip=]{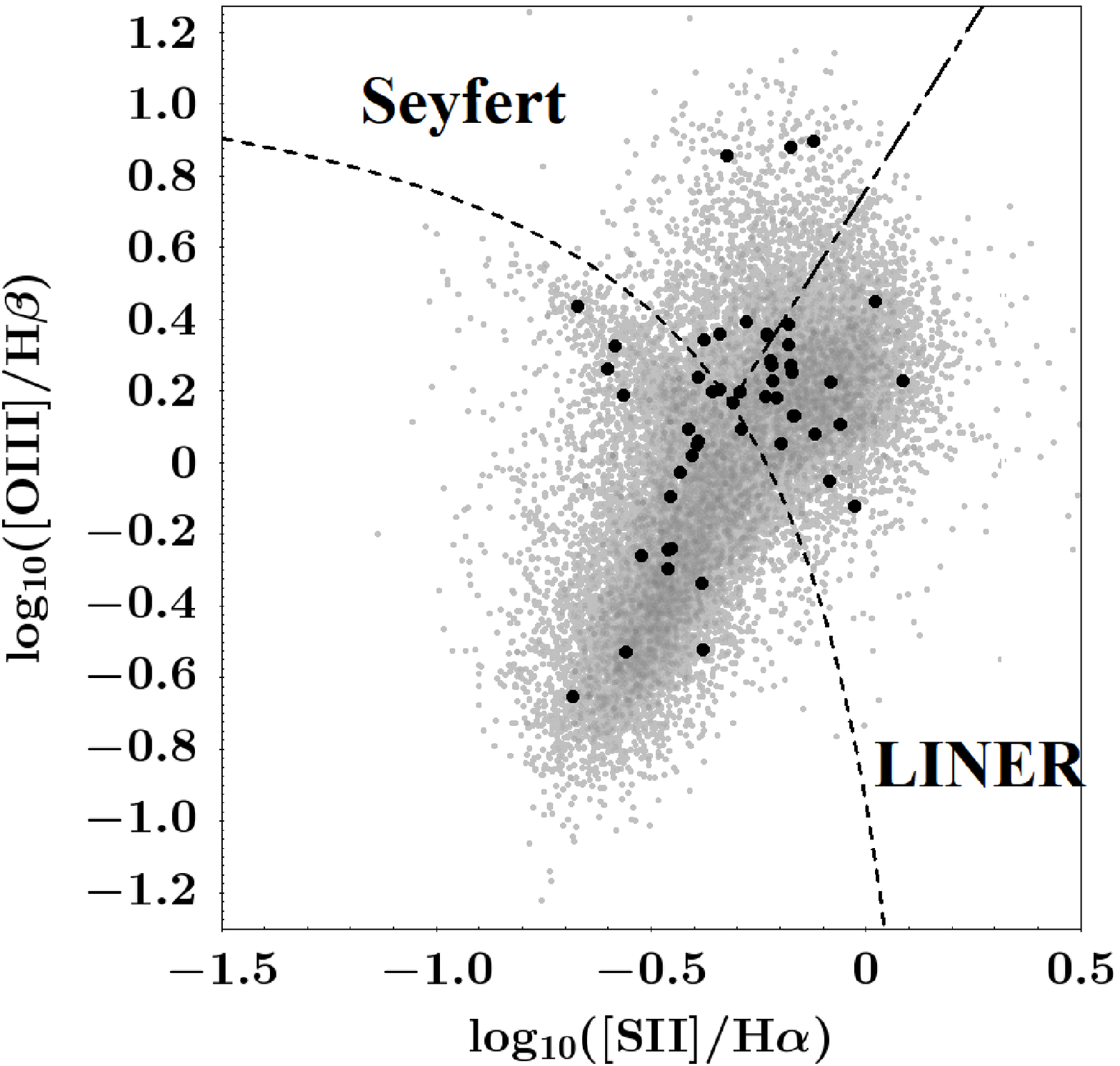}\\

\vspace{0.4cm}

\includegraphics[width=7.5cm, clip=]{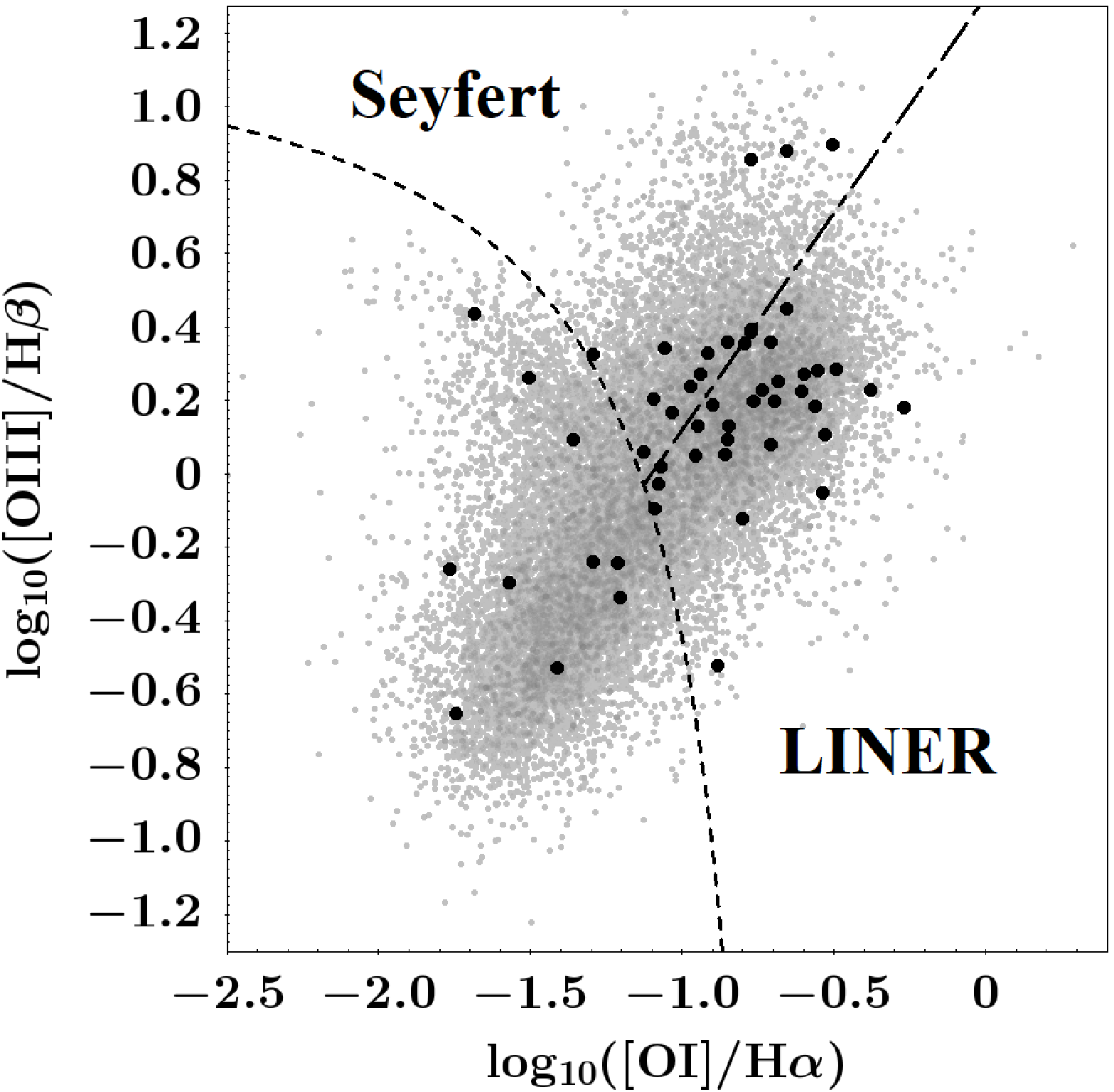}\\

\vspace{0.4cm}

\caption{BPT diagrams for the best PRG candidates (black dots) and galaxies 
from SDSS (gray dots). The lines indicate the boundaries for different 
types of AGNs according to Kewley et al. (2006).}
%\label{fig2}
\end{figure}
%%%%%%%%%%%%%%%%%%%%%%%%%%%%%%%%%%%%%%%%%%%%%%%%%%%%%%%%%%%%%%%%%%%%

Figure\,2 shows the diagnostic BPT diagrams for the best PRG candidates 
(including the objects from RM2019) and the comparison sample from SDSS.
The dashed lines in the figure indicate the lines separating the regions 
of star-forming galaxies (they lie below these lines) from the AGNs 
(they are located, respectively, above these lines). The line segments
on the diagrams indicate the boundary lines for the separation of AGNs into 
Seyfert galaxies (SyG) and LINERs (Low-Ionization Narrow Emission-line 
Region). We attributed the galaxies with composite nuclei, whose spectra 
carry signatures of ionization by both an active nucleus and young stars, 
to AGNs, as is occasionally done (see, e.g., Hwang et al. 2012;
Sabater et al. 2013, 2015; Kim et al. 2020). On the one hand, this was done 
to increase the statistics. On the other hand, on diagnostic diagrams of a 
different type the composite nuclei can fall into the AGN region
(see, e.g., Trouille et al. 2011).

The characteristics of the demarcation lines were taken from Kewley et al. (2006). 
The BPT diagrams for all of the galaxy samples listed in Table\,1 were
constructed in a similar way. In what follows, we assume that a galaxy has 
an active nucleus only if it was attributed to AGNs on all three diagnostic
diagrams.

%%%%%%%%%%%%%%%%%%%%%%%%%%%%%%%%%%%%%%%%%%%%%%%%%%%%%%%%%%%%%%%%%%%%
\section{Results and discussion}
%%%%%%%%%%%%%%%%%%%%%%%%%%%%%%%%%%%%%%%%%%%%%%%%%%%%%%%%%%%%%%%%%%%%

The results of our analysis are summarized in Table\,2. 
Columns 2 and 3 in this table provide the AGN
fraction (in percent) for a given PRG subgroup (the
number of AGNs in this subgroup is given in parentheses) and the 
analogous fraction for the galaxies of the corresponding comparison 
sample. The numbers for the comparison samples were obtained by averaging 
over 50 independent realizations of these samples.
Because of the large size of the comparison samples,
the standard errors of the fractions are small ($\sim$0.1\%)
and are not given in Table\,2.

\begin{table}
\caption{The fraction of AGNs among PRGs and galaxies from SDSS}
\begin{center}
\begin{tabular}{c|c|c}
\hline
Sample           & PRGs & Comparison   \\
                  &     & sample  \\
\hline               
All PRGs           & 26\% $\pm$ 3\% (48) &  25\%      \\
Best candidates + RM2019 & 50\% $\pm$ 7\% (26) &  40\%      \\
Good candidates & 12\% $\pm$ 4\% (10) &  16\%      \\
Related objects & 24\% $\pm$ 8\% (7)  &  29\%      \\
Face-on rings   & 24\% $\pm$ 9\% (5)  &  26\%       \\   
\hline
\end{tabular}
\end{center}
\end{table}

It can be seen from Table\,2 that the fractions of AGNs for the PRG 
candidates and the comparison samples are comparable. Only the best candidates
show a slight excess of the AGN fraction; among them half of the galaxies 
have active nuclei, while they account for about 40\% in the comparison 
sample. Previously, a close estimate was obtained for the objects from the 
catalogue by Whitmore et al. (1990), though based on a smaller PRG sample and with
less homogeneous spectroscopic data (Reshetnikov et al. 2001).

Consider the statistics of AGNs among the best PRG candidates in more detail. 
According to our classification, 6, 16, and 4 galaxies among them
can be attributed to SyG (11.5\% $\pm$ 4.4\%), LINERs (30.8\% $\pm$ 6.4\%), 
and galaxies with composite nuclei (7.7\% $\pm$ 3.7\%), respectively. 
The corresponding numbers for the comparison sample are 6.9\%, 23.0\%,
and 9.8\%. Hence it can be seen that within our poor statistics the fractions 
of composite nuclei in both samples are comparable, while the number of SyG
and LINERs among PRGs is increased. Note also that the ratio of the number of 
LINERs to the number of Seyfert galaxies among PRGs is $\approx$2.7, which is
close to the corresponding ratio in the comparison sample ($\approx$3.3). 
Approximately the same ratio of LINERs and SyG is also obtained by other authors,
although it can depend on various factors, including, in particular, the galaxy 
mass (see, e.g., Kewley et al. 2006; Sabater et al. 2013).

Figure\,3 shows the dependences of the fractions of AGNs of different types 
on the integrated luminosity of the galaxies. A well-known observational
trend is clearly seen in the figure -- the AGN fraction increases with galaxy 
luminosity (and, accordingly, mass) (Kauffmann et al. 2003). This trend is traced
both for the objects from the comparison sample and for all of the galaxies 
from SDSS considered by us. It can be seen from Fig.\,3 that at $M_r \leq -20^m$ 
there is an approximately twofold excess of the AGN fractions
among PRGs compared to normal galaxies of the same luminosity. Similarly, we 
compared the fraction of AGNs as a function of the galaxy $g - r$ color.
The results turned out to be less clear than those for their luminosities, 
but, on the whole, for normal galaxies there is a tendency for the AGN fraction 
to increase with $g - r$. As with the luminosities, AGNs are encountered in 
PRGs with $g - r \approx 1$ relatively more frequently than in normal galaxies 
with the same color.

%%%%%%%%%%%%%%%%%%%%%%%%%%%%%%%%%%%%%%%%%%%%%%%%%%%%%%%%%%%%%%%%%%%%
\begin{figure}
\centering
\includegraphics[width=7cm, angle=-90, clip=]{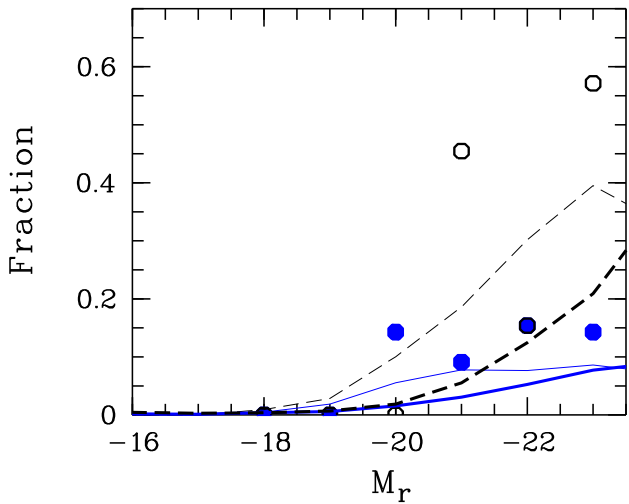}\\
\caption{Fraction of Seyfert galaxies (filled circles) and LINERs (open circles) 
versus $r$-band absolute magnitude of the galaxy among the best PRG candidates. 
The solid lines indicate the dependences for SyG based on the comparison sample 
(thin line) and for the entire sample of galaxies from SDSS (thick line). 
The dashed lines indicate the dependences for LINERs based on the
comparison sample (thin line) and the sample of all SDSS galaxies (thick line).}
%\label{fig3}
\end{figure}
%%%%%%%%%%%%%%%%%%%%%%%%%%%%%%%%%%%%%%%%%%%%%%%%%%%%%%%%%%%%%%%%%%%%

Note that our results are based on small PRGs statistics and are preliminary. 
The conclusion about an increased fraction of AGNs among PRGs is 
supported by the fact that an excess of Seyfert galaxies and LINERs among 
PRGs is traced over almost all of the independently considered luminosity 
intervals at $M_r \leq -20^m$ (Fig.\,3).

A simpler way to identify AGNs among PRGs is to use the original classification 
in SDSS. According to SDSS data, 9 galaxies among the best PRG candidates 
are classified as ``AGN'' or ``QSO'', giving an AGN fraction of 17$\pm$5\%. 
In the comparison sample there are fewer such galaxies, only 6.8\% $\pm$ 0.2\%. 
Note that the spectral classification in SDSS is less detailed than ours 
and there is no separation into Seyfert galaxies and LINERs in this survey.

\begin{table*}
\caption{Candidates for Seyfert galaxies among PRGs}
\begin{center}
\begin{tabular}{c|c|c|c|c|c|c}
\hline
Galaxy         & SDSS & $M_r$  & $g-r$ & $\Delta P.A.$         & $D_r$ & $D_r/D_h$ \\
               & type &        &       & ($^{\circ}$)& (kpc) &  \\
\hline               
SPRC-1            & QSO AGN & -22.01 & 0.80  & 82 & 16 & 0.83         \\
SPRC-90           & AGN     & -21.55 & 0.71  & 86 & 37 & 1.53         \\
SPRC-156          & AGN     & -20.66 & 0.73  & 73 & 19 & 0.80         \\
SPRC-161          & QSO     & -20.54 & 0.80  & 84 & 15 & 1.53      \\
SPRC-258          & AGN     & -21.83 & 0.61  & -- & 30 & 1.42         \\
$N$14 (RM2019)  & AGN     & -20.86 & 0.75  & 85 & 18 & 1.07        \\
\hline
\end{tabular}
\end{center}
\end{table*}

Table\,3 lists the main characteristics of 6 galaxies from the SPRC and RM2019, 
which, according to our data, can be attributed to SyG, and, at the same
time, they are classified as AGNs in SDSS as well. Columns 1, 2, 3, and 4 in 
the table give, respectively, the galaxy name according to the SPRC or the 
object number according to Table\,1 in RM2019, the spectral classification 
in SDSS, the absolute magnitude, and the color (Aguado et al. 2019) 
corrected for Galactic extinction and the $k$-correction (Schlafly
and Finkbeiner 2011; Chilingarian et al. 2010). The next three columns 
summarize the characteristics of the polar structures found by us from 
$r$-band images: the observed angle between the major axis of the
central galaxy and the polar structure ($\Delta$P.A.), the diameter ($D_r$), 
and the ratio of the polar structure diameter to the central galaxy diameter 
($D_r/D_h$). (For SPRC-258, which belongs to the objects with
a presumably nearly face-on polar ring, $\Delta$P.A. was not measured.) 
The data in Table\,3 show that the PRGs with active nuclei are similar in 
their characteristics to the normal bright PRGs (see Table\,2 in
RM2019). We can only note the small relative size of the polar structures 
($\langle D_r/D_h \rangle  = 1.20 \pm 0.34$) and their almost orthogonal 
orientation ($90^{\circ} - \Delta P.A. = 8^{\circ} \pm 5^{\circ}$). Most of 
the objects from Table\,3 exhibit dust lanes associated with the polar rings, 
suggesting the presence of gas in them.

The question of what can stimulate the nonthermal activity of galactic nuclei 
(the local and global environment, the gravitational interaction and mergers
of galaxies, their secular evolution, etc.) still remains the subject of debates 
(see, e.g., Sabater et al. 2013) and, therefore, we will restrict ourselves 
only to a few remarks.

A necessary, but not sufficient condition for the triggering of nuclear activity 
is the presence of a cold gas in the inner regions of the galaxies. 
The environment of the galaxies, their interaction and mergers
affect the nuclear activity indirectly, supplying the gas to the galactic 
centers; however, other mechanisms are apparently responsible for the transport 
of gas to the immediate neighborhood of the black hole (Reichard et al. 2009; 
Sabater et al. 2015).

In most PRGs a gas-poor main galaxy is surrounded by a polar structure that, 
as a rule, is gas-rich (van Grokom et al. 1987; van Driel et al. 2000).
It can be assumed that the interaction of the central galaxy with this structure 
leads to the gas infall onto the central galaxy. Evidence of the current 
interaction between the polar-ring gas and the central galaxy have recently 
been found in a number of PRGs (Egorov and Moiseev 2019). The small
relative size of the polar structures in the galaxies from Table\,3 is 
consistent with the assumption about the formation of shocks at their inner 
boundaries (Wakamatsu 1993) and the subsequent accretion of matter onto the 
galaxies. The subsequent behavior of this gas and the mechanisms of angular 
momentum loss by its remain unexplored.

Objects related to PRGs are the so-called galaxies with {\it inner polar structures}
(for a review, see Moiseev 2012). A stellar-gaseous disk $\sim$1 kpc in size
almost orthogonal to the plane of the main galaxy is present in the central 
regions of such objects. The inner polar structures are encountered more often
than the outer large-scale polar rings. According to Sil'chenko (2016), they 
are observed in $\sim$10\% of the S0 galaxies in the local Universe. As in the
case of PRGs, the origin of the inner structures is associated with the 
external accretion of matter, for example, with the capture and disruption of 
a gas-rich dwarf companion (Sil'chenko et al. 2011). Thus, in the case of PRGs, 
there is an external reservoir of matter accreting in a circumpolar plane onto the
central galaxy. In the case of inner polar structures, such a reservoir formed 
by the captured matter is observed directly in the circumnuclear region of the
galaxies. It can be assumed that many AGNs can also be observed among galaxies 
with inner polar structures. An examination of NED\footnote{NASA/IPAC
Extragalactic Database -- http://ned.ipac.caltech.edu} showed that
about a third of the galaxies from the list by Moiseev (2012) are classified 
by this database as having active nuclei, including such known Seyfert galaxies
as NGC\,1068 and NGC\,3227. Unfortunately, the list of known galaxies with 
inner polar structures is so far very inhomogeneous and partly random. 
Increasing the number of such objects and obtaining homogeneous spectroscopic 
data for them will allow, in particular, the possible connection between the
formation of inner polar structures and the nuclear activity of galaxies to be 
considered.

%%%%%%%%%%%%%%%%%%%%%%%%%%%%%%%%%%%%%%%%%%%%%%%%%%%%%%%%%%%%%%%%%%%%
\section{Conclusions}
%%%%%%%%%%%%%%%%%%%%%%%%%%%%%%%%%%%%%%%%%%%%%%%%%%%%%%%%%%%%%%%%%%%%

Based on SDSS spectra, we considered the frequency of AGNs among galaxies 
with large-scale optical polar structures. An analysis of the BPT diagrams
and the classification of the types of nuclei published in SDSS led to the 
conclusion about a possible excess of AGNs among the best PRG candidates. This
conclusion is based on an as yet relatively small number of known PRGs and 
it needs a further confirmation.

A detailed study and numerical simulations of the interaction of the polar 
structures (both outer and inner ones) with the central galaxies can give useful
information about the triggering mechanisms of non-thermal nuclear activity 
in galaxies.

%%%%%%%%%%%%%%%%%%%%%%%%%%%%%%%%%%%%%%%%%%%%%%%%%%%%%%%%%%%%%%%%%%%%
\section{Acknowledgments}
%%%%%%%%%%%%%%%%%%%%%%%%%%%%%%%%%%%%%%%%%%%%%%%%%%%%%%%%%%%%%%%%%%%%

We are grateful to the referees for the helpful remarks
that improved the presentation of our results.

We used the NASA/IPAC Extragalactic Database (NED), which is operated by
the Jet Propulsion Laboratory, the California Institute of Technology, under 
contract with the National Aeronautics and Space Administration (USA).

This study is based on public SDSS data. SDSS-IV is funded by the Alfred P. Sloan 
Foundation, the U.S. Department of Energy Office of Science, and the Partici-
pating Institutions. SDSS-IV acknowledges support and resources from the Center 
for High-Performance Computing at the University of Utah. The SDSS web site is
www.sdss.org.

SDSS is managed by the Astrophysical Research Consortium for the Participating 
Institutions of the SDSS Collaboration including the Brazilian Participation Group,
the Carnegie Institution, the Carnegie Mellon University, the Chilean Participation 
Group, the French Participation Group, the Harvard-Smithsonian Center for 
Astrophysics, the Instituto de Astrofisica de Canarias, the Johns
Hopkins University, the Kavli Institute for the Physics and Mathematics of the 
Universe (IPMU)/University of Tokyo, the Korean Participation Group, the Lawrence
Berkeley National Laboratory, the Leibniz Institute for Astrophysics Potsdam (AIP), 
the Max Planck Institute for Astronomy (MPIA), the Max Planck Institute for
Astrophysics (MPA), the Max Planck Institute for Extraterrestrial Physics (MPE), 
the Chinese National Astronomical Observatory, the New Mexico State University,
the New York University, the Notre Dame University, the National Observatory/MCTI, 
the Ohio State University, the Pennsylvania State University, the Shanghai 
Astronomical Observatory, the United Kingdom Participation Group, the Mexican 
National University, the University of Arizona,the University of Colorado in 
Boulder, the Oxford University, the University of Portsmouth, the University
of Utah, the University of Virginia, the University of Washington, the University 
of Wisconsin, the Vanderbilt University, and the Yale University.

%\pagebreak

%%%%%%%%%%%%%%%%%%%%%%%%%%%%%%%%%%%%%%%%%%%%%%%%%%%%%%%%%%%%%%%%%%%%
\begin{center}
{\Large \bf References}
\end{center}
%%%%%%%%%%%%%%%%%%%%%%%%%%%%%%%%%%%%%%%%%%%%%%%%%%%%%%%%%%%%%%%%%%%%

\small

\noindent
D.S. Aguado, R. Ahumada, A. Almeida, et al., Astrophys. J. Suppl. Ser. 240, 
23 (2019). \\

\noindent
J.A. Baldwin, M.M. Phillips, and R. Terlevich, Publ.
Astron. Soc. Pacif. 93, 5 (1981). \\

\noindent
K. Bekki, Astrophys. J. 490, L37 (1997). \\

\noindent
F. Bournaud and F. Combes, Astron. Astrophys. 401,
817 (2003). \\

\noindent
Ch.B. Brook, F. Governato, Th. Quinn, J. Wadsley,
A.M. Brooks, B. Willman, A. Stilp, and P. Jonsson,
Astrophys. J. 689, 678 (2008). \\

\noindent
D. Calzetti, AIP Conf. Proc. 408, 403 (1997). \\

\noindent
I. Chilingarian, A.-L. Melchior, and I. Zolotukhin,
Mon. Not. R. Astron. Soc. 405, 1409 (2010). \\

\noindent
O.V. Egorov and A.V. Moiseev, Mon. Not. R. Astron.
Soc. 486, 4186 (2019). \\

\noindent
I. Finkelman, J.G. Funes, and N. Brosch, Mon. Not.
R. Astron. Soc. 422, 2386 (2012). \\

\noindent
J.H. van Grokom, P.L. Schechter, and J. Kristian,
Astrophys. J. 314, 457 (1987). \\

\noindent
H.S. Hwang, C. Park, D. Elbaz, and Y.-Y. Choi,
Astron. Astrophys. 538, A15 (2012). \\

\noindent
G. Kauffmann, T. M. Heckman, Ch. Tremoni, J. Brinchmann, S. Charlot, 
S.D.M. White, S.E. Ridgway, J. Brinkmann, et al., Mon. Not. R.
Astron. Soc. 346, 1055 (2003). \\

\noindent
L.J. Kewley, M.A. Dopita, R.S. Sutherland,
C.A. Heisler, and J. Trevena, Astrophys. J. 556, 121 (2001). \\

\noindent
L.J. Kewley, B. Groves, G. Kauffmann, and T. Heckman, 
Mon. Not. R. Astron. Soc. 372, 961 (2006). \\

\noindent
M. Kim, Y.-Y. Choi, and S.S. Kim, Mon. Not. R.
Astron. Soc. 491, 4045 (2020). \\

\noindent
A.V. Maccio, B. Moore, and J. Stadel, Astrophys. J.
636, L25 (2006). \\

\noindent
A.V. Moiseev, Astrophys. Bull. 67, 147 (2012). \\

\noindent
A.V. Moiseev, K.I. Smirnova, A.A. Smirnova, and
V.P. Reshetnikov, Mon. Not. R. Astron. Soc. 418, 244 (2011). \\

\noindent
A. Moiseev, S. Khoperskov, A. Khoperskov, K. Smirnova, A. Smirnova, 
A. Saburova, and V. Reshetnikov, Baltic Astron. 24, 76 (2015). \\

\noindent
D.E. Osterbrock and G.J. Ferland, Astrophysics of Gaseous Nebulae and Active 
Galactic Nuclei, 2nd ed. (Univ. Sci. Books, Sausalito, CA, 2006). \\

\noindent
T.A. Reichard, T.M. Heckman, G. Rudnick,
J. Brinchmann, G. Kauffmann, and V. Wild, Astrophys. J. 691, 1005 (2009). \\

\noindent
V.P. Reshetnikov and F. Combes, Astron. Astrophys. 291, 57 (1994). \\

\noindent
V. Reshetnikov and F. Combes, Mon. Not. R. Astron. Soc. 447, 2287 (2015). \\

\noindent
V.P. Reshetnikov and A.V. Mosenkov, Mon. Not. R.
Astron. Soc. 483, 1470 (2019). \\

\noindent
V. Reshetnikov and N. Sotnikova, Astron. Astrophys. 325, 933 (1997). \\

\noindent
V.P. Reshetnikov, M. Fa\'undez-Abans,  and  M. de Oliveira-Abans, 
Mon. Not. R. Astron. Soc. 322, 689 (2001). \\

\noindent
V.P. Reshetnikov, M. Fa\'undez-Abans, and
M. de Oliveira-Abans, Astron. Lett. 37, 171 (2011). \\

\noindent
J. Sabater, P.N. Best, and M. Argudo-Fern\'andez,
Mon. Not. R. Astron. Soc. 430, 638 (2013). \\

\noindent
J. Sabater, P.N. Best, and T. M. Heckman, Mon. Not.
R. Astron. Soc. 447, 110 (2015). \\
 
\noindent 
S.S. Savchenko and V.P. Reshetnikov, Astron. Lett. 43, 146 (2017). \\
 
\noindent
E.F. Schlafly and D.P. Finkbeiner, Astrophys. J. 737, 103 (2011). \\

\noindent
O.K. Sil'chenko, Astron. J. 152, 73 (2016). \\

\noindent
O.K. Sil'chenko, I.V. Chilingarian, N.Ya. Sotnikova,
and V.L. Afanasiev, Mon. Not. R. Astron. Soc. 414,  3645 (2011). \\

\noindent
M.B. Taylor, ASP Conf. Ser. 347, 29 (2005). \\

\noindent
L. Trouille, A.J. Barger, and C. Tremoni, Astrophys.
J. 742, 46 (2011). \\

\noindent
W. van Driel, M. Arnaboldi, F. Combes, and
L. S. Sparke, Astron. Astrophys. Suppl. 141, 385 (2000). \\

\noindent 
S. Veilleux and D.E. Osterbrock, Astrophys.  J. Suppl. Ser. 63, 295 (1987). \\

\noindent
K.-I. Wakamatsu, Astron. J. 105, 1745 (1993). \\

\noindent
B.C. Whitmore, D.B. McElroy, and F. Schweizer,
Astrophys. J. 314, 439 (1987). \\

\noindent 
B.C. Whitmore, R.A. Lucas, D.B. McElroy, T.Y. Steinman-Cameron, 
P.D. Sackett, and  R.P. Olling, Astron. J. 100, 1489 (1990). \\

%%%%%%%%%%%%%%%%%%%%%%%%%%%%%%%%%%%%%%%%%%%%%%%%%%%%%%%%%%%%%%%%%%%%
\end{document}